\documentstyle[12pt, multicol, psfig]{article}

\newcommand{\tr}{\mathop{\rm tr}\nolimits}

\newcommand{\vr}{{\bf r}}

\newcommand{\D}{{\cal D}}
\newcommand{\F}{{\cal F}}

\begin{document}

\author{D.E.Khmelnitskii\\
{\it Cavendish Laboratory, University of Cambridge}\\ {\it Cambridge CB3 0HE, UK }\\ {\it and}\\ {\it Institut Laue-Langevin, B.P. 156}\\ {\it F-38042 Grenoble, Cedex 9, FRANCE}}

\title{\bf ON ANDERSON LOCALISATION\\ 
 OF TWO-DIMENSIONAL\\ ELECTRONS \\ 
IN WEAK MAGNETIC FIELD}  

\date{1 March, 2005}

\maketitle

\begin{abstract}
The two-parameter RG flow diagramme \cite{DK83}  
is used for obtaining the magnetic field dependence of localisation length 
$L_c(B)$ for charged particles in 2D random potential at low disorder and 
weak magnetic fields $B$. The result reproduces the periodic singularities 
of $L_c(B)$ . 
\end{abstract}

\section*{\center Introduction}
This short note represents a further development of the old paper by
this author \cite{DK84} (see also the comment \cite{La84}). A prediction 
has been made in Ref \cite{DK84} and  \cite{La84} that the charged particles 
in a two-dimensional random potential and weak perpendicular magnetic field 
$B$ experience delocalisation at the energy corresponding their chemical 
potential $E_F$ and particular values of the field.  It had not been noticed 
then that the same two-parameter RG flow diagramme of Ref
\cite{DK83}, which is the basis of the conclusion in Ref
\cite{DK84}, allows to obtain a general form of the field dependence 
of localisation length $L_c(B)$.

\section*{\center Two-parameter Scaling}

The Ref \cite{DK83} was based upon the field theory \cite{PR83}  
for two-dimensional electrons in the random potential and magnetic field. 
The partition function $Z$ of this field theory could be expresses through 
the functional integral over matrices $ Q$, subject to the constraint 
$Q^2 = 1$ :
\begin{equation}
\label{Pr} 
Z = \int_{Q^2 =1} \D Q ~e^{-\F} ; \, \,  \F = \frac{\pi \hbar}{4 e^2} \tr \int d\vr 
\left\{ \sigma_{xx} \left(\nabla Q \right)^2 + 2 \sigma_{xy} ~Q 
\left[\partial_x Q, \partial_y Q \right]\right\}  
\end{equation}
where the coefficients $\sigma_{xx}$ and $\sigma_{xy}$ in the free energy 
$\F$ correspond to the values of the dissipative and Hall conductivities at 
the sample size of the order of the mean free path $l$. The term at the front 
of the Hall conductivity  $\sigma_{xy}$ is a topologic invariant proportional to an integer number. Therefore the partition function $Z$ does not change if the transformation 
$$ \sigma_{xy} \to \sigma_{xy} + \frac{e^2}{2 \pi \hbar} $$ 
is made. Integration over fast variations of $Q(\vr)$ leads to renormalization group equations 
 \begin{eqnarray}
\label{RG1} 
&&\frac{d \ln \sigma_{xx}}{\ln L} = \beta_{xx} \left(\sigma_{xx} ,  \sigma_{xy}  \right),\\
&&\frac{d \ln \sigma_{xy}}{\ln L} = \beta_{xy} \left(\sigma_{xx} ,  \sigma_{xy}  \right)
 \label{RG2} 
\end{eqnarray}
We know several basic properties of scaling functions $\beta_{xx} \left(\sigma_{xx} ,  \sigma_{xy}  \right)$ and $\beta_{xy} \left(\sigma_{xx} ,  \sigma_{xy}  \right)$:
\begin{itemize}
\item At $\sigma_{xx} \gg e^2/\hbar$ the function $\beta_{xx} \left(\sigma_{xx} , \sigma_{xy}   \right)$ does not depend on its second argument; 
\item both functions $\beta_{xx} \left(\sigma_{xx} ,  \sigma_{xy}  \right)$ and $\beta_{xy} \left(\sigma_{xx} ,  \sigma_{xy}  \right)$ are periodic functions of their second argument with period $e^2/2\pi\hbar$;
\item  The function $\beta_{xx} \left(\sigma_{xx} ,  0  \right)$ is negative at all $\sigma_{xx} > 0$
\item  The function $\beta_{xx} \left(\sigma_{xx} ,  e^2/ 4 \pi \hbar  \right)$ has a fixed point at $\sigma_{xx} = \sigma_* \sim e^2/ \hbar$ \cite{LLP84}.
\item The function $\beta_{xx}$ is the even function of  $\sigma_{xy}$, while 
the function $\beta_{xy}$ is its odd function. Therefore, $\beta_{xy}$ vanishes at  $\sigma_{xy}$ equal to semi-integer of $e^2/2\pi \hbar$  
\end{itemize}
Combining these basic facts, one can conjecture the flow diagramme of the system (\ref{RG1}, \ref{RG2}) , shown in Figure 1, as it has been done in Ref\cite{DK84}. 
\begin{figure}
\protect
\label{Fig 1}
\centerline{\psfig{figure=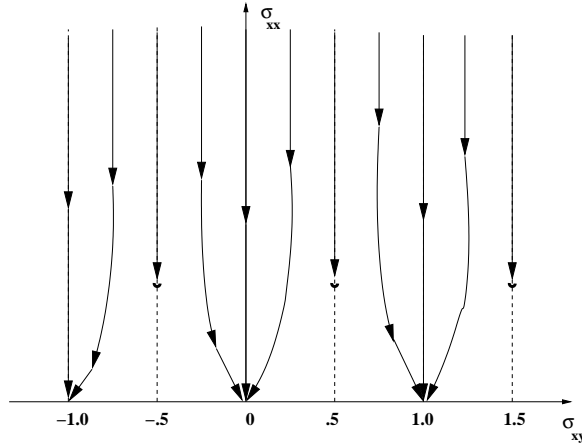,width=3.00in} }
\vskip 20pt
\caption{\rm The RG Flow Diagramme for Two-parameter Scaling for Localisation of a Quantum Particle in 2D Random Potential and Magnetic Field}
\end{figure}
It has been shown in Ref \cite{DK84} that, if the long-distance cut-off 
monotonically increases with the temperature decrease, then the flow 
diagramme reproduces qualitatively the Integer Quantum Hall Phenomenology. 

\section*{\center ``Levitation'' of Extended States}

The  RG-equations (\ref{RG1},  \ref{RG2}) with  the flow  diagramme of
Figure  1  allows to  treat  Anderson  Localisation  in the  limit  of
classically  weak  magnetic  magnetic   fields  $\Omega  \tau  \ll  1$
\cite{DK84},\cite{La84}, where $\Omega  = e B / m  c$ is the cyclotron
frequency and  $\tau$ is  transport mean free  time. In this  limit the
dissipative conductivity  at short distances  $\sigma_{xx}$ depends on
magnetic field weakly and could be considered a constant
\begin{equation}
\sigma_{xx} = \frac{e^2}{2 \pi \hbar} ~g, \qquad g \gg 1. 
\label{xx}
\end{equation}
As for Hall conductivity  $\sigma_{xy}$ , 
\begin{equation}
\sigma_{xy} = \Omega \tau ~\sigma_{xx} = \frac{e^2}{2 \pi \hbar} ~g ~\Omega \tau 
\label{xy}
\end{equation}
According to the flow diagramme, all states of electrons are localised
but those for which $\sigma_{xy}$, given by Eq~(\ref{xy}), equals to
odd semi-integer $e^2/ 2 \pi \hbar$. Therefore, if $\sigma_{xy} = (e^2/
4 \pi \hbar) ( 2 n + 1) $, the level of the chemical potential
$E_F$ of electrons coincides with the $n$-th delocalised state. Since 
 $\sigma_{xy}$ decreases with decreasing the fields $B$, the energy $E_n$ of 
 the $n$-th delocalised state increases as 
\begin{equation}
E_n = \frac{\hbar}{\Omega \tau^2 } ~\left(n + \frac{1}{2} \right) 
\label{lev}
\end{equation}
This prediction is called ``levitation\footnote{This word is suggested by R.B.Laughlin}  of extended states'' with decreasing of magnetic field.  

\section*{\center New Result}

The point, overlooked in previous publications, is that the same flow 
diagramme allows to find out dependence of localisation length $L_c$ on  
magnetic field $B$. If $g \gg 1$ the functions $\beta_{xy}$ are small at 
almost all relevant values of $\sigma_{xx}$ but at the very end of 
renormalization. This leads to factorization of dependences of localisation 
length $L_c$ on  $\sigma_{xx}$ and 
 $\sigma_{xy}$. As the result 
\begin{equation} 
L_c (B)  \approx 
l  ~F \left( \Omega ~\tau ~g \right) \exp\left[\frac{\pi^2}{2}~g^2 \right], 
\label{answer}
\end{equation}
where $F(x) \ge 1$, universal periodic function with period one. This 
function is equal to unity $( F(x) = 1 )$ at all integer values of the 
argument $x$ and has an infinite singularity at semi-integer values.  

\section*{\center Conclusion}

Magnetic field dependence of localisation length $L_c(B)$ in Eq~(\ref{answer}) strongly resembles dependence of the gap in one-dimensional metal as 
a function of the number of electrons per unit cell \cite{LS}, despite 
the RG in 1D problem is very different. It remains to be found whether 
this similarity means existence of an actual mapping of one of the problem 
on the other or exhibits a coincidence.

\section*{\center Acknowledgment}

The author is grateful to E.I.~Kats and P.~Nozieres for hospitality at ILL and   encouraging discussions on the subject of this note.



\begin{thebibliography}{gg}
%
\bibitem{DK83} D.E.~Khmelnitskii {\it JETP Letters} {\bf 38 }, 454    (1983)
\bibitem{DK84}  D.E.~Khmelnitskii {\it Phys. Lett. A} {\bf 106A}, 182 (1984)
%
\bibitem{La84} R.B.~Laughlin ~{\it Phys.Rev.Lett.} {\bf 52}, 2304 (1984)
%
\bibitem{PR83} A.M.M.~Pruisken {\it Nucl. Phys.} {\bf B235}, 277 (1984)
%
\bibitem{LLP84} H.~Levine, S.B.~Libby and  A.M.M.~Pruisken {\it Phys.Rev.Lett.} {\bf 51}, 1915 (1983)
%
\bibitem{LS} A.I.Larkin and J.Sak {\it Phys. Rev. Lett.}{\bf 39 }, 1026 (1977) 
%
\end{thebibliography}
\end{document}